\documentclass[prd, aps, superscriptaddress, preprintnumbers, twocolumn, floatfix, nofootinbib]{revtex4}
\pdfoutput=1

\usepackage{amsfonts}
\usepackage{amsmath}
\usepackage{amssymb}
\usepackage{bm}
\usepackage{dcolumn}
\usepackage{graphicx}   
\usepackage[latin1]{inputenc}
\usepackage{latexsym}
\usepackage{rotating}
\usepackage{hyperref}
\usepackage{graphicx}
\usepackage{color}

\newcommand\be{\begin{equation}}
\newcommand\ba{\begin{eqnarray}}
\newcommand\ee{\end{equation}}
\newcommand\ea{\end{eqnarray}}

\newcommand{\cA}{{\cal{A}}}

\begin{document}

\title{Moduli and Graviton Production during Moduli Stabilization}

\author{Mesbah Alsarraj}
\email{mesbah.alsarraj@mail.mcgill.ca}
\affiliation{Department of Physics, McGill University, Montr\'{e}al, QC, H3A 2T8, Canada} 
 
\author{Robert Brandenberger}
\email{rhb@physics.mcgill.ca}
\affiliation{Department of Physics, McGill University, Montr\'{e}al, QC, H3A 2T8, Canada}

\date{\today}

\begin{abstract}

In theories beyond the Standard Model, in particular in theories with extra spatial dimensions such as superstring theory, there are a large number of scalar fields which appear in the low energy effective action. These moduli fields must be stabilized. Often, moduli stabilization involves a stage during which the moduli fields oscillate coherently about their ground state value. Here, we study moduli and graviton production during the period during which the background modulus field is oscillating, assuming that this period takes place in the radiation phase. We find a resonant production of moduli fluctuations, and a tachyonic instability to the generation of long wavelength gravitational waves. As a consequence, the period of moduli stabilization is short on a Hubble time scale.

\end{abstract}

\maketitle

\section{Introduction} 
\label{sec:intro}

In theories beyond the Standard Model of particle physics, in particular in models with extra spatial dimensions such as superstring theory, there are many scalar fields which appear in the low energy effective action. These are called {\it moduli fields}. Examples include the size and shape moduli of the extra spatial dimensions, the moduli describing the position of branes, and the dilaton itself (see e.g. \cite{Baumann} for a discussion of a number of moduli fields which can play a role in early universe cosmology). In order to be consistent with experimental constraints, all of these moduli fields need to be stabilized at minima of their respective effective potentials. 

There is a useful analogy with the cosmology of the QCD axion which is the angular component of a complex scalar field (see e.g. \cite{axion} for a review of the cosmology of the axion). At high temperatures, the potential for the axion field vanishes, and the axion can take on any value. However, at low temperatures a non-perturbative potential for the axion is generated, and the axion configuration will relax to the minimum of this potential. If the axion field is initially homogeneously distributed over our Hubble patch (which is a result of early universe models such as Inflation \cite{Guth} or the Ekpyrotic scenario \cite{Ekp}), then the initial stage of axion stabilization will involve a period during which the axion field background undergoes damped oscillations about its vacuum value. This process happens during the radiation phase of Standard cosmology.

In analogy, the situation we have in mind is one where the modulus field starts to homogeneously oscillate about the minimum of an effective potential which develops after some phase transition. We assume that the phase transition happens during the radiation phase of Standard cosmology. Our analysis is independent of that the theory is which describes the very early universe, which could be Inflation, a bouncing scenario such as Ekpyrosis, or an emergent scenario such as String Gas cosmology \cite{BV}.

As we show here, the oscillations of the modulus field induce oscillatory correction terms to the cosmological scale factor. These correction terms, in turn, appear in the equations of motion for both moduli fluctuations and graviton perturbations. We demonstrate that the new terms result in an instability for moduli production which is similar to the parametric resonance instability at the onset of the reheating period after inflation \cite{DK, TB} (see \cite{RBrev, Karouby} for reviews of this preheating instability). We also show that the dominant effect in the equation of motion for gravitons is a tachyonic instability for long wavelength modes, analogous to the tachyonic preheating instability \cite{tachyonic} which appears in some models of inflation. These instabilities lead to a rapid (on the Hubble time scale) transfer of energy from the initial modulus condensate to a gas of modulus and graviton fluctuations. The modulus fluctuations carry more energy than the gravitons.

The outline of this paper is as follows. In the next section we study the effects which the oscillating moduli background field has on the cosmological scale factor. In Section 3 we then study the induced production of moduli field fluctuations, and in Section 4 we analyze the production of gravitational waves. Section 5 focuses on a computation of the energy transfer from the initial homogeneous modulus condensate to a gas of fluctuations of moduli and gravitons, and on an estimate of when back-reaction shuts off the resonant particle production process. We end with a discussion of our results.

We will use units in which the speed of light, Boltzmann's constant and Planck's constant are set to $1$. We work in the context of a spatially flat Friedmann universe with metric
\be
ds^2 \, = \, dt^2 - a(t)^2 d{\bf x}^2 \, \equiv \, a(t(\eta))^2 \bigl( d\eta^2 - d{\bf x}^2 \bigr) \, ,
\ee
where $t$ is physical time, and ${\bf x}$ are the comoving spatial coordinates. Often, the equations simplify if we make use of conformal time $\eta$. The scale factor is $a(t)$, and during the radiation phase of Standard Big Bang cosmology which we consider here, the background scale factor is given by
\be \label{scaleF}
a(t) \, = \, \bigl( \frac{t}{t_0} \bigr)^{1/2} \, ,
\ee
where $t_0$ is a normalization constant which we will take to be the time when the modulus field starts to roll. The Hubble rate is given by
\be
H(t) \, \equiv \, \frac{\dot{a}}{a} \, ,
\ee
where an overdot stands for the derivative with respect to time. Newton's gravitational constant is denoted, as usual, by $G$, and it determines the Planck mass $m_{pl}$ via $G \equiv m_{pl}^{-2}$. We will denote the derivative with respect to conformal time by a prime.

\section{Moduli-Induced Corrections to the Scale Factor Evolution} \label{review}
  
We consider a modulus field $\varphi$ whose low energy effective potential takes the form
\be
V(\varphi) \, = \, \frac{1}{2} m^2 \varphi^2 \, ,
\ee
where $m$ is the mass of the modulus field. Unless there are special symmetries which suppress the potential, the above is a fairly general assumption for the potential of a modulus field. We assume that $\varphi$ is trapped at a value $\varphi = {\cal{A}}$ at high temperatures, and that it starts to roll at a time $t_0$ which we take to be during the epoch of radiation-domination of Standard Big Bang cosmology.

The equation for the homogeneous modulus background field is
\be
{\ddot{\varphi}} + 3H {\dot{\varphi}} + m^2 \varphi^2 \, = \, 0 \, ,
\ee
neglecting the couplings of the modulus field to other matter fields. We introduced a rescaled field $\psi$ via
\be \label{scaling}
\varphi \, = \, a^{-3/2} \psi \, .
\ee
In terms of the rescaled field, the equation of motion becomes
\be
{\ddot{\psi}} - \bigl(\frac{9}{4} H^2 - \frac{3}{2} {\dot{H}} \bigr) \psi + m^2 \psi \, = \, 0 \, .
\ee
We will assume that the modulus field oscillates rapidly on a Hubble time scale (and we will show later on that this leads to self-consistent calculations). In this case, the terms in the above equation proportional to $H^2$ and ${\dot{H}}$ can be dropped, and we obtain an approximate solution
\be \label{osc}
\psi(t) \, = \, \cA {\rm{cos}} (m (t - t_0)) \, .
\ee
This oscillating modulus field will induce oscillations in the scale factor superimposed on the background scale factor $a_0(t)$.

The matter content of the universe can be modelled by a superposition of radiation with an equation of state $w_r \equiv p_r / \rho_r = 1/3$ (where $p_r$ and $\rho_r$ are radiation pressure and energy density, respectively), and the modulus field $\varphi$ whose energy density and pressure are given by
\ba
\rho_{\varphi} \, &=& \frac{1}{2} {\dot{\varphi}}^2 + V(\varphi) \\
p_{\varphi} \, &=& \frac{1}{2} {\dot{\varphi}}^2 - V(\varphi) \, . \nonumber
\ea 

We make the following ansatz for the scale factor
\be \label{corr}
a(\eta) \, \equiv \, a_0(\eta) + b(\eta) \, ,
\ee
where $b(\eta)$ is the correction due to the presence of the oscillating modulus field. We will work under the assumption that the scalar field yields a small correction to the energy density, and that hence $b(\eta) \ll a(\eta)$, and we will work to first order in an expansion in the small quantity $b(\eta)$.

The dynamical equation of motion for the scale factor $a(\eta)$ is
\be \label{FRW}
\frac{a^{\prime \prime}}{a} \, = \, \frac{4 \pi G}{3} a^2 \bigl( \rho - 3 p \bigr) \, .
\ee
For pure radiation the right hand side of this equation vanishes, and hence, using (\ref{scaleF}), (\ref{FRW}) becomes
\be
b^{\prime \prime} \, = \, \frac{4 \pi G}{3} \bigl( a_0 + b \bigr)^3 \bigl( \rho_{\varphi} - 3 p_{\varphi} \bigr) \, .
\ee
To lowest order in $b$ this becomes
\be \label{beq}
b^{\prime \prime} \, = \, \frac{4 \pi G}{3} a_0^3 \bigl( \rho_{\varphi} - 3 p_{\varphi} \bigr) \, .
\ee

Since the scale factor in the radiation epoch is (\ref{scaleF})
and since physical time and conformal time are related by
\be
\eta(t) \, = \, 2 \bigl( t t_0 \bigr)^{1/2} \, ,
\ee
Equation (\ref{beq}) becomes
\be \label{beq3}
b^{\prime \prime} \, = \, \frac{4 \pi G}{3} \frac{1}{8}
\frac{\eta^3}{t_0^3} \bigl( \rho_{\varphi} - 3 p_{\varphi} \bigr) \, .
\ee

This equation can be solved using the Green's function method
\be \label{GFsol}
b(\eta) \, = \, \int_{\eta_0}^{\eta} d\eta^{\prime} G(\eta - \eta^{\prime}) s(\eta^{\prime}) \, ,
\ee
where $s(\eta^{\prime})$ is the right hand side (the source) of Eq. (\ref{beq3}) and $G(\eta - \eta^{\prime})$ is the Green's function of the above equation, i.e. the solution of
\be
G^{\prime \prime} (\eta - \eta^{\prime}) \, = \, \delta(\eta - \eta^{\prime}) \, ,
\ee
which vanishes for $\eta < \eta^{\prime}$. The Green's function is
\be
G(\eta - \eta^{\prime}) \, = \, \eta - \eta^{\prime} \, ,
\ee
and hence (\ref{GFsol}) becomes
\be \label{GFsol2}
b(\eta) \, = \, \int_{\eta_0}^{\eta} d\eta^{\prime} (\eta - \eta^{\prime}) s(\eta^{\prime}) \, .
\ee

In the approximation that the time scale of oscillation is smaller than the Hubble time scale, i.e. $ mt \gg 1$, we can neglect the time derivative of the scale factor when working out the form of the source term $s(\eta^{\prime})$ and, using (\ref{osc}) and (\ref{scaling}) we obtain
\be
s(\eta^{\prime}) \, = \frac{4 \pi}{3} G \bigl[2m^2\cA^2 - 3m^2\cA^2{\rm{sin}}^2(m(t - t_0)) \bigr]
\, .
\ee
The second term can be written as a constant plus a term which oscillates at twice the frequency. In evaluating the Green's function integral (\ref{GFsol2}), the integral over the constant terms in $s(\eta^{\prime})$ give a contribution to $b(\eta)$ which is proportional to $(\eta - \eta_0)^2$, while the integral over the term containing the oscillatory part of $s(\eta^{\prime})$ can be approximated by using
\be
\int_0^x x^{\prime} {\cal{O}}(x^{\prime}) dx^{\prime} \, \simeq \frac{x}{\omega} {\cal{O}}(x) \, ,
\ee
where ${\cal{O}}$ is the oscillatory function and $\omega$ is its frequency of oscillation. With this approximation, we obtain
\ba \label{beq2}
b(\eta) \, &\simeq& \, \frac{\pi}{3} G m^2 \cA^2 \bigl(\eta - \eta_0 \bigr)^2 \\
& & + \pi G m^2 \cA^2 \frac{\eta - \eta_0}{m} {\rm{cos}}[\frac{1}{2} \frac{m}{t_0} (\eta^2 - \eta_0^2)] \nonumber
\ea

In the following section we will study the effects which this correction to the scale factor has on the equation of motion for the moduli fluctuations, and in  Section 4 we turn to the investigation of the consequences for graviton production.

\section{Resonant Moduli Production} \label{analysis}

The equation of motion for the modulus field in the quadratic potential we are considering is linear, and hence all Fourier modes evolve independently. Thus, we Fourier expand the field in terms of comoving coordinates, with $k$ denoting the wavenumber. After the field rescaling (\ref{scaling}), the equation for the k'th Fourier mode becomes
\be \label{psikeq}
{\ddot{\psi}}_k + \bigl(m^2 - \frac{3}{2} {\dot{H}} - \frac{9}{4} H^2 \bigr) \psi_k + \frac{k^2}{a^2} \psi_k \, = \, 0 \, .
\ee
In the following we will consider long wavelength modes with $k/a(t_0) < m$, and we will hence neglect the gradient term in this equation. Note that in this section it is more convenient to work in terms of physical time instead of conformal time.

The main point of the following discussion is that the oscillatory correction $b(\eta)$ to the scale factor induces oscillatory terms in the equation of motion (\ref{psikeq}) for the Fourier modes. The equation then becomes of Floquet type, and we expect a parametric resonance instability \cite{Landau, Arnold}. Thus, we expect resonant production of moduli fluctuations (see \cite{Natalia} for an early study of this process).

To extract the leading contribution of $b(\eta)$ in the above equation, we insert the expansion of the scale factor (\ref{corr}) into the gravitational contribution to the mass term in the above equation (\ref{psikeq}) and expand to linear order in $b$, resulting in
\ba \label{squeeze}
\frac{9}{4} H^2 + \frac{3}{2} {\dot{H}} \, &=& \, \frac{3}{2} \frac{\ddot{a_0}}{a_0} + \frac{3}{4} H_0^2 \\
&+& \frac{3}{2} H_0 \frac{\dot{b}}{a_0} - \frac{3}{2} H_0^2 \frac{b}{a_0} 
- \frac{3}{2} \frac{\ddot{a_0}}{a_0} \frac{b}{a_0} + \frac{3}{2} \frac{\ddot{b}}{a_0} \, . \nonumber
\ea 

Inserting the result (\ref{beq2}) for $b(\eta)$ we obtain an expression containing many terms. Of all of the terms linear in $b$ and its derivatives, we will extract the leading term in the ``rapid oscillation'' approximation $m t_0 \gg 1$. The leading term comes from the ${\ddot{b}}$ term in (\ref{squeeze}), and within this term the leading contribution comes from the term involving the second derivative of the oscillating term in $b$. Thus, with these approximations the leading contribution to the gravitational contribution to the mass is
\ba \label{squeeze2}
\frac{9}{4} H^2 + \frac{3}{2} {\dot{H}} \, &\simeq& \, \frac{3}{2} \frac{\ddot{a_0}}{a_0} + \frac{3}{4} H_0^2 \\
&+& \, \frac{3}{2} \pi G m^3 \cA^2 (\eta - \eta_0) \frac{\eta^2}{t t_0 a_0} 
{\rm{cos}}[\frac{1}{2} \frac{m}{t_0} (\eta^2 - \eta_0^2)] \, . \nonumber
\ea
It is the term in the second line in (\ref{squeeze2}) which is responsible for the resonant production of $\psi_k$ particles.

As was done in the case of inflationary preheating in \cite{TB}, we will assume that particle production is rapid on the Hubble time scale. In this case we can neglect the terms involving time derivatives of $a_0$, and we can set $a_0 = 1$. The equation of motion (\ref{psikeq}) then becomes (recall that we are considering infrared modes with $k < m$)
\be \label{psikeq2}
{\ddot{\psi}}_k + \bigl(m^2 - \frac{3}{2} \pi G m^3 \cA^2 8 (t - t_0) 
{\rm{cos}}[2 m (t - t_0)]) \psi_k  \, = \, 0 \, ,
\ee
where we have expressed the conformal time $\eta$ in terms of physical time $t$. Apart from the fact that the correction term to the mass also has an amplitude which is increasing in time, this equation has the form of the Mathieu equation \cite{Landau, Arnold} and leads to a parametric resonance instability. 

To obtain an analytical estimate of the resulting particle production, we will replace the factor $(t - t_0)$ by its order of magnitude $t_0$. In parallel, we have numerically solved the equation of motion without this approximation, and we will present the numerical results below. For the analytical treatment, we introduce a re-scaled dimensionless time $z = mt$, and we will here (only in this subsection) denote the derivative with respect to $z$ by an overdot. The equation then becomes
\be
{\ddot{\psi_k}} + \bigl( 1 - q {\rm{cos}}[2z] \bigr) \psi_k \, = \, 0 \, ,
\ee
with
\be
q \, = \, 12 \pi G \cA^2 m t_0 \, .
\ee
As long as
\be
\frac{\cA}{m_{pl}} \, > \frac{1}{2 \sqrt{3}} \bigl( m t_0 \bigr)^{-1/2}
\ee
we are in the {\it broad resonance} regime \cite{RBrev}, and the solutions will be of the form
\be \label{psisol}
\psi_k \, \sim \, \psi_k(t_0) e^{\mu (t - t_0)} {\rm{cos}}[2mt] \, ,
\ee
with Floquet exponent $\mu$ given by
\be \label{Floq}
\mu \, \simeq \, \sqrt{q} \, = \, \frac{\cA}{m_{pl}} (12 m t_0)^{1/2} m \, .
\ee
Note that when the solution is written in terms of the dimensionless time $z$, the Floquet exponent is dimensionless and does not contain the final factor of $m$ in (\ref{Floq}). Note also that the Floquet exponent is independent of $k$ (for the infrared modes which we are considering), and that the dimensionless exponent is predicted to scale as $m^{1/2}$.

Since the analytical study makes use of a number of approximations, it is important to solve the mode equation (\ref{psikeq}) numerically in order to verify that the analytical approximations are not missing essential physics. Thus, we have solved (\ref{psikeq}) numerically using Mathematica. The result is shown in Fig. 1 and shows good agreement with what is obtained using the analytical approximation, showing oscillations with an exponentially increasing amplitude.
 
 \begin{figure}[h!]
\includegraphics[width=\hsize]{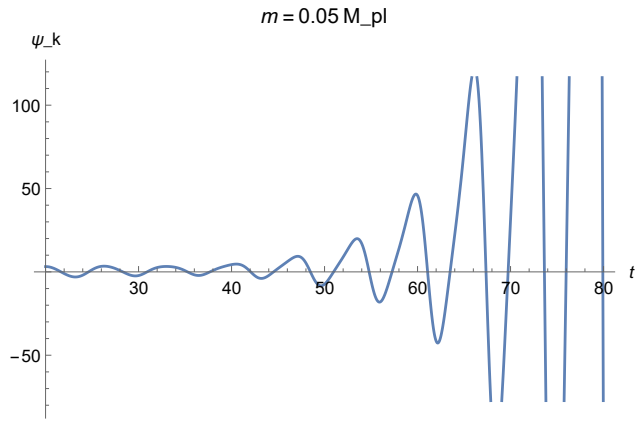}
\caption{ Time evolution of the fluctuation field $\psi_k$. Dimensionless units are used in which the time coordinate (horizontal axis) and the field mode (vertical axis) are in Planck units.  The initial value of the field was taken to be given by the vacuum initial values for a massive scalar field, i.e. $\psi_k(t_0) = 1 / \sqrt{2m}$.} 
\end{figure}

In Figure 2, the numerical results for the dimensionless Floquet exponent (the exponent given by (\ref{Floq}) without the final factor of $m$) as a function of mass $m$ is shown. For small values of the mass, the increase of the Floquet exponent agrees reasonably well with the analytical prediction. For larger values of $m$, however, a saturation of the Floquet exponent is observed.

\begin{figure}[h!]
\includegraphics[width=\hsize]{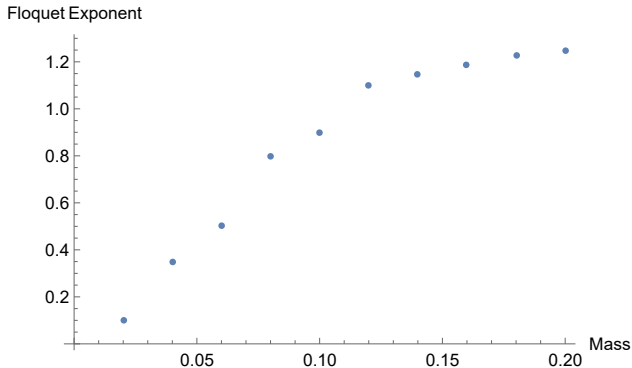}
\caption{The dimensionless Floquet exponent $\mu$ (vertical axis) as a function of mass $m$ (horizontal axis, in Planck units). } 
\end{figure}

Having determined the value of the Floquet exponent, we can study under which conditions the modulus production process is rapid in a Hubble time scale. The condition for this to be the case is
\be
\mu t_0 \, \gg \, 1 \, ,
\ee
which, making use of (\ref{Floq}), becomes
\be
\frac{\cA}{m_{pl}} 2 \sqrt{3} \bigl( m t_0 \bigr)^{3/2} \, .
\ee
Assuming that the modulus condensate begins to roll when the modulus energy density equals the background density yields
\be
t_0 \, = \, \bigl( \frac{3}{32\pi} \bigr)^{1/2} \frac{m_{pl}}{\cA} m^{-1} \, .
\ee
This yields
\be
\mu t_0 \, = \, \frac{9}{64\pi} \frac{1}{\sqrt{2\pi}} \frac{m_{pl}}{\cA} \, ,
\ee
and we conclude that provided 
\be \label{cond}
\cA \, \ll \, m_{pl} \, 
\ee
then the energy transfer is rapid.

Note that if the moduli fluctuations originate at $t_0$ in their vacuum state, the spectral shape remains the same since the Floquet exponent is independent (for infrared modes) of $k$. Hence, in this case the induced fluctuations will have a negligible effect on the curvature perturbations, independent of how the modulus field is coupled to the dominant radiation field. The fluctuations of $\varphi$ represent entropy perturbations, and these can in principle seed curvature fluctuations (see \cite{MFB, Gordon} for reviews, and \cite{Minos} for an early study in the case of axion perturbations)). If, on the other hand, the modulus fluctuations at $t_0$ inherit a scale-invariant spectrum from a previous early universe phase such as Inflation or Ekpyrosis, then the mechanism described in this paper will boost this spectrum and make it possible for the modulus fluctuations to impact curvature perturbations on length scales relevant to cosmological observations. However, the results will be very model-dependent (see e.g. \cite{Hossein} for some studies), and we will not pursue this avenue of investigation here.

\section{Graviton Production}

The ansatz for the metric including gravitational waves in an expanding universe in a spatially flat Friedmann-Robertson-Walker-Lemaitre universe is 
\be
ds^2 \, = \, a^2(\eta) [d\eta^2 - (\delta_{ij} - h_{ij})dx_i dx^j \, ,
\ee
where $i$ and $j$ run over the spatial indices. The gravitational wave tensor $h_{ij}$ is transverse and tracelss and can be decomposed into two polarization states. In linear theory, these states and each Fourier mode thereof evolve independently. Let $h_k$ denote the amplitude of the k'th Fourier mode of a given polarization state. In terms of the rescaled field 
\be
\mu_k \, \equiv \, a h_k G^{-1/2} \, ,
\ee
where the factor $G^{-1/2}$ is introduced such that the field $\mu$ has the dimensions of mass like a regular canonical scalar field, the equation of motion \footnote{As is usual in the theory of cosmological perturbations, it is more convenient here to work in terms of conformal time.} is \cite{MFB, RHBfluctsrev}
\be \label{mueq}
\mu_k^{\prime \prime} + \bigl( k^2 - \frac{a^{\prime \prime}}{a} \bigr) \mu_k \, = \, 0 \, .
\ee
Since in a radiation-dominated background $a_0^{\prime \prime} = 0$, then to leading order in the amplitude of the correction $b(\eta)$ to the scale factor, this mode equation becomes
\be
\mu_k^{\prime \prime} + \bigl( k^2 - \frac{b^{\prime \prime}}{a_0} \bigr) \mu_k \, = \, 0 \, .
\ee

Making use of the expression for $b(\eta)$ derived in Section 2 (see (\ref{beq2})) we have (where in the oscillatory terms we have only kept the terms which dominate in the case $mt_0 \gg 1$ which we are considering)
\ba \label{squeezing}
b^{\prime \prime} \, &=& \, \frac{2 \pi}{3} G m^2 \cA^2 \\ 
&-& \pi G m^2 \cA^2 \bigl( \frac{\eta}{t_0} \bigr)^2 m (\eta - \eta_0) 
{\rm{cos}}[\frac{m}{2t_0} (\eta^2 - \eta_0^2 )] \nonumber
\ea
As is apparent by inserting (\ref{squeezing}) into the equation (\ref{mueq}) for $\mu_k$, the variation of $b(\eta)$ leads to instabilities for long wavelength modes. The constant term in $b^{\prime \prime}$ sources a tachyonic instability, similar to the tachyonic resonance instability which arises in some models of inflationary reheating \cite{tachyonic}. The oscillatory term will induce further instabilities of parametric resonance type, similar to the ones we studied for dilaton fluctuations in the previous section.

Like in the previous section we will assume that the instabilities occur on a time scale small compared to the Hubble time scale. Hence, we can neglect the time dependence of $a_0$ and set $a_0 = 1$. The equation for the tachyonic resonance is
\be
\mu_k^{\prime \prime} + \bigl( k^2 - \frac{2 \pi}{3}Gm^2\cA^2 \bigr) \mu_k \, = \, 0 \, .
\ee
As is evident, there is a critical wavenumber $k_c$ given by
\be \label{kcrit}
k_c \, = \, \sqrt{\frac{2\pi}{3}} \frac{\cA}{m_{pl}} m \, ,
\ee
and for $k < k_c$ the amplitude of $\mu_k$ grows exponentially
\be
\mu_k \, \sim \, e^{k_c(\eta - \eta_0)} \, .
\ee

Inserting the values for $k_c$ and for $\eta - \eta_0 \sim \eta_0$ we can check that the time scale of the instability is comparable to the Hubble time scale but not significantly smaller, as it should be to fully justify the approximations we have made. Hence, the analytical analysis in this section will be less reliable than they were in the case of moduli production. This comparison shows that the energy transfer is mostly into moduli fluctuations.

If the gravitons are initially in their vacuum state, then with the resulting initial condition 
\be
\mu_k(\eta_0) \, = \, \frac{1}{\sqrt{2k}} 
\ee
(quantum vacuum normalization of a canonical scalar field), then the induced energy density $\rho_{gw}$ in gravitons is
\ba \label{gwenergy}
\rho_{gw} \, &\sim& \,  4\pi \int_0^{k_c} dk k^2 k^2 \mu_k(\eta_0)^2 e^{2 k_c(\eta - \eta_0)} \nonumber \\
&\sim & \, k_c^4 e^{2 k_c(\eta - \eta_0)}
\ea

Note that since the growth rate of the Fourier modes is independent of $k$ (for infrared modes), the induced spectrum of gravitational waves retains its vacuum shape (by shape we refer to the k-dependence). If the initial conditions for the gravitons at the time $\eta_0$ correspond to an excited state with a scale-invariant spectrum (as is predicted in various early universe models, e.g. in Inflation \cite{Starob} or in String Gas cosmology \cite{NBPV}), then the final spectrum retains its scale-invariant shape, but with a boosted amplitude. The effect discussed here thus will increase the predicted tensor to scalar ratio, but, as discussed in the following section, the increase in amplitude will not be large.

\section{Energy Transfer and Back-Reaction}

Moduli and graviton production terminates once back-reaction effects become important in the dynamics. This will occur when the energy density in the produced moduli fields or in the gravitons (whichever occurs earlier) becomes comparable to the initial energy density in $\varphi$. Above, we have shown that the energy transfer into moduli fields dominates over graviton production.

In the case when moduli production is rapid on the Hubble time scale, i.e. when the condition (\ref{cond}) is satisfied, the energy density $\rho_{\varphi}$ in moduli fluctuations can be estimated by
\ba \label{energy}
\rho_{\varphi} \, &\sim& \, 4\pi \int_o^m dk k^2 k^2 e^{2\mu(t - t_0)} \frac{1}{2k} \, , \nonumber \\
&=& \, \frac{\pi}{2} m^4 e^{2\mu(t - t_0)} \, .
\ea
In the first line, the first factor of $k^2$ inside the integral comes from the phase space of modes, the second one from the gradients (we are computing the gradient energy), and the final factor comes from assuming that the initial conditions for $\psi_k$ are given by quantum vacuum perturbations. Since the instability we have studied only holds for modes with $k < m$, the integral is cut off at $k = m$.

Back-reaction will shut off the instability once the energy density in the fluctuations is comparable to the initial energy density in the modulus condensate, i.e. when
\be
\rho_{\varphi} \, \sim \, \frac{1}{2} m^2 \cA^2 \, .
\ee
From (\ref{energy}), it follows that the resulting time period $\tau$ of the instability is
\be \label{period}
\tau \, = \, \frac{1}{2} \mu^{-1} {\rm{ln}}[ \frac{1}{\pi} \bigl( \frac{\cA}{m} \bigr)^2] \, .
\ee
We see that the energy transfer is rapid on the initial Hubble time scale.

We can now use this result to estimate the energy density which flows into gravitational waves. Starting with (\ref{gwenergy}), expressing conformal time in terms of physical time and expanding the exponent to leading order in the instability period $\tau$ we find
\be
\rho_{gw} \, \sim \, k_c^4 e^{4 k_c \tau} \, .
\ee
Inserting the values of $k_c$ from (\ref{kcrit}) and $\tau$ from (\ref{period}) we find that the exponent is typically smaller than $1$. Thus the ratios of gravitational wave and moduli fluctuation energies at the end of the resonant period is
\be
\frac{\rho_{gw}}{\rho{\varphi}} \, \sim \, \frac{k_c^4}{m^2 \cA^2} \, 
\sim \, \frac{\cA^2 m^2}{m_{pl}^4} \, .
\ee

In the previous section we pointed out that graviton production will increase the predicted tensor to scalar ratio. However, because the total energy transfer into gravitons is not large, the increase in the ratio is not large.
\\

\section{Conclusions and Discussion} \label{conclusion}

Theories beyond the Standard Model have many scalar moduli fields which need to be stabilized. Here, we  have studied the production of moduli fluctuations and gravitons during a process in which an initial homogeneous modulus field is stabilized while this condensate is oscillating about the ground state, assuming that this process happens in the radiation phase of Standard Big Bang cosmology.  We found that moduli production is very efficient, taking place on a time scale which is short compared to the Hubble time scale. The production process is a parametric resonance instability. Gravitons are also produced via an instability, in this case a tachyonic instability. In both cases, it is infrared modes which are amplified. The amplification of the moduli fluctuations is more efficient then that of gravitons.

From the point of view of the theory of cosmological perturbations, modulus fields are entropy mode. Thus, the process we have studied corresponds to the parametric amplification of entropy fluctuations, set up in the primordial universe, during the period of radiation domination. Depending on the coupling of the moduli to the radiation field, these entropy fluctuations may seed an important contribution to curvature fluctuations.

We have assumed that the modulus field gives a subdominant contribution to the total energy density, thus justifying the perturbative treatment we have given. Our analysis could easily be generatised to the case of a background which is not dominated by radiation - it is only the form of the background about which we expand which would change. For example, our analysis could easily be generalized to a matter phase in the very early universe, e.g. during inflationary reheating. The instabilities which we study are similar to the instabilities by which the inflaton field decays into particles during reheating. 

Our analysis applies to moduli fields predicted by superstring theory such as the dilaton or K\''ahler and complex structure moduli. It also applies to axionic fields which are displaced from their low temperature ground state in the early universe. Applied to axions, our results imply that the axion field very quickly loses its coherence on length scales larger than its Compton wavelength.

\section*{Acknowledgement}

\noindent RB wishes to thank the Pauli Center and the Institutes of Theoretical Physics and of Particle- and Astrophysics of the ETH for hospitality. The research of RB at McGill is supported in part by funds from NSERC and from the Canada Research Chair program.  MA would like to acknowledge the Abdullah Al Ghurair Foundation for Education (AGFE) and the Office of Graduate and Post-Doctoral Studies at McGill University for funding, and in addition Heliudson Bernardo for useful discussions.

\end{document}